\documentclass[conference]{IEEEtran}
\IEEEoverridecommandlockouts
\usepackage{cite}
\usepackage{amsmath,amssymb,amsfonts}
\usepackage{mathtools}
\usepackage{graphicx}
\usepackage{textcomp}
\usepackage{xcolor}
\usepackage{hyperref}
\usepackage{booktabs} % for thicker lines
\usepackage{placeins}  % For \FloatBarrier
\usepackage{threeparttable}
\usepackage{float}
\hypersetup{
    colorlinks=true,
    linkcolor=black,      % color of internal links
    citecolor=black,      % color of citations
    filecolor=black,   % color of file links
    urlcolor=black        % color of external links
}
\def\BibTeX{{\rm B\kern-.05em{\sc i\kern-.025em b}\kern-.08em
    T\kern-.1667em\lower.7ex\hbox{E}\kern-.125emX}}

\usepackage{algpseudocode}
\usepackage{algorithm}
\usepackage{amsmath}

\begin{document}

\title{\Large \textbf{Augmentation of EEG and ECG Time Series for Deep Learning Applications: Integrating Changepoint Detection into the iAAFT Surrogates}
\thanks{Nina Moutonnet was supported by the UKRI Centre for Doctoral Training in Artificial Intelligence for Healthcare (EP/S023283/1).}
%\thanks{Identify applicable funding agency here. Identify applicable funding agency here.}
}

\author{
    \IEEEauthorblockN{Nina Moutonnet$^{1}$, Gregory Scott$^{2}$, Danilo P. Mandic$^{3}$}
    \IEEEauthorblockA{$^{1}$Department of Computing, Imperial College London, UK \\
    $^{2}$Department of Brain Sciences, Imperial College London, UK \\
    $^{3}$Department of Electrical and Electronic Engineering, Imperial College London, UK \\
    Email: nm2318@imperial.ac.uk, gregory.scott99@imperial.ac.uk, d.mandic@imperial.ac.uk}
}

\maketitle
\begin{abstract}

The performance of deep learning methods critically depends on the quality and quantity of the available training data. This is especially the case for physiological time series, which are both noisy and scarce, which calls for data augmentation to artificially increase the size of datasets. Another issue is that the time-evolving statistical properties of nonstationary signals prevent the use of standard data augmentation techniques. To this end, we introduce a novel method for augmenting nonstationary time series. This is achieved by combining offline changepoint detection with the iterative amplitude-adjusted Fourier transform (iAAFT), which ensures that the time-frequency properties of the original signal are preserved during augmentation. The proposed method is validated through comparisons of the performance of i) a deep learning seizure detection algorithm on both the original and augmented versions of the CHB-MIT and Siena scalp electroencephalography (EEG) databases, and ii) a deep learning atrial fibrillation (AF) detection algorithm on the original and augmented versions of the Computing in Cardiology Challenge 2017 dataset. By virtue of the proposed method, for the CHB-MIT and Siena datasets respectively, accuracy rose by 4.4\% and 1.9\%, precision by 10\% and 5.5\%, recall by 3.6\% and 0.9\%, and F1 by 4.2\% and 1.4\%. For the AF classification task, accuracy rose by 0.3\%, precision by 2.1\%, recall by 0.8\%, and F1 by 2.1\%.
\end{abstract}


\section{\bfseries Introduction}

The success of deep learning in the analysis of time series has also highlighted that large amount of training data is required to avoid overfitting and ensure generalisability. However, obtaining high quality labelled data can be challenging in real-world applications. In particular, medical data is often laborious and/or dangerous to collect, poorly labelled, and associated with data privacy concerns. Data augmentation is an effective way to artificially increase the size of existing datasets by creating synthetic copies which retain the statistical properties of the original data \cite{augmentation, fawaz}. Whilst this has been extensively studied in computer vision, data augmentation for time series, such as in electroencephalography (EEG) and electrocardiography (ECG), is still in its infancy. In addition, temporal data augmentation is challenging due to the typically nonlinear and nonstationary nature of the data, lack of an intuitive understanding of the data, and unknown signal generating mechanisms. The lack of high-quality surrogate data is therefore an obstacle towards automatic analysis of EEG and ECG recordings using machine learning (ML), as open-source labelled datasets are still quite limited in size \cite{review}. \\

Current data augmentation strategies range from simple time-domain techniques through to generative models such as time-GAN \cite{timeGAN}. These include time series manipulation, such as flipping along the vertical and horizontal axes \cite{augmentation}, downsampling \cite{data_aug}, and Gaussian noise injection \cite{noise}. More advanced techniques include dynamic time warping \cite{DTW}, the iterative amplitude-adjusted Fourier transform (iAAFT)\cite{data_aug}, injecting complex patterns to a signal such as spikes and trends \cite{data_aug2, review_dataug}, and generative AI models such as time-GAN and variational auto-encoders \cite{AE, ecg_aug}. 

However, the choice of a data augmentation method that preserves the relevant characteristics of the original time series is not straightforward. For example, EEG and ECG recordings usually have a close to zero mean and do not exhibit long-term trends, but rather spikes and waves. Consequently, most simple signal augmenting manipulation techniques are not appropriate for EEG and ECG time series, e.g. time reversing a signal is only applicable to time series that are symmetric about the horizontal axis, which is not the case with EEG and ECG data. Similarly, while the iAAFT surrogates have equivalent first and second-order statistics to the original data \cite{iaaft}, they operate only on stationary data and are unsuitable for data augmentation of nonstationary time series. Generative AI models, despite their elegance, are extremely difficult to tune, and lack robustness as nonstationary data do not have a fixed probability density function. More specifically, the heterogeneity and sparsity of data poses a significant challenge when using DL-based augmentation strategies, a characteristic inherent to both ECG and EEG recordings. These physiological signals exhibit notable inter-patient variability, and their pathological manifestations encompass a broad spectrum of abnormalities. In the context of EEG, seizures may manifest as either generalised disturbances, present across all the channels, or as focal abnormalities localised to specific cortical regions and thus only visible in a subset of channels \cite{review}. Similarly, cardiac arrhythmias, a type of atrial fibrillation (AF), demonstrate considerable heterogeneity in their electrophysiological expression, characterised by irregular R-R intervals and irregular morphological alterations in the PQRST complex \cite{arr}. \\

To this end, the aim of this work is to help mitigate these issues by incorporating changepoint detection into data augmentation strategies. For EEG, this is achieved by assuming that the signal is composed of piecewise quasi-stationary segments. The existing nonstationary surrogates require prior knowledge of the changepoints of dynamical regimes in data, which is rarely available in practice \cite{CP_review, data_aug2}. The proposed data augmentation method, on the other hand, employs a changepoint detection scheme to segment the time series and applies a modified version of the iAAFT algorithm to each segment individually, which makes it suitable to nonstationary data. Experimental results in the context of deep learning for epileptic EEG support the analysis. Unlike the EEG augmentation approach, signal integrity during ECG augmentation is preserved through peak detection and retention rather than temporal segmentation. The peaks remain invariant during iAAFT augmentation, facilitating convergence to a surrogate that conserves the fundamental ECG morphology — and thus potentially significant clinical information — whilst introducing controlled variability in non-peak regions. 

\section{\bfseries Proposed algorithm}

The proposed algorithm addresses nonstationarity through a two-stage process: initial detection of either quasi-stationary segment boundaries in EEG or characteristic peaks in ECG, followed by the use of the modified iAAFT algorithm for surrogate data generation.

\subsection{Changepoint detection for EEG}\label{changepointsection}

\begin{figure*}[!ht]
    \centering
    \includegraphics[width=0.9\textwidth]{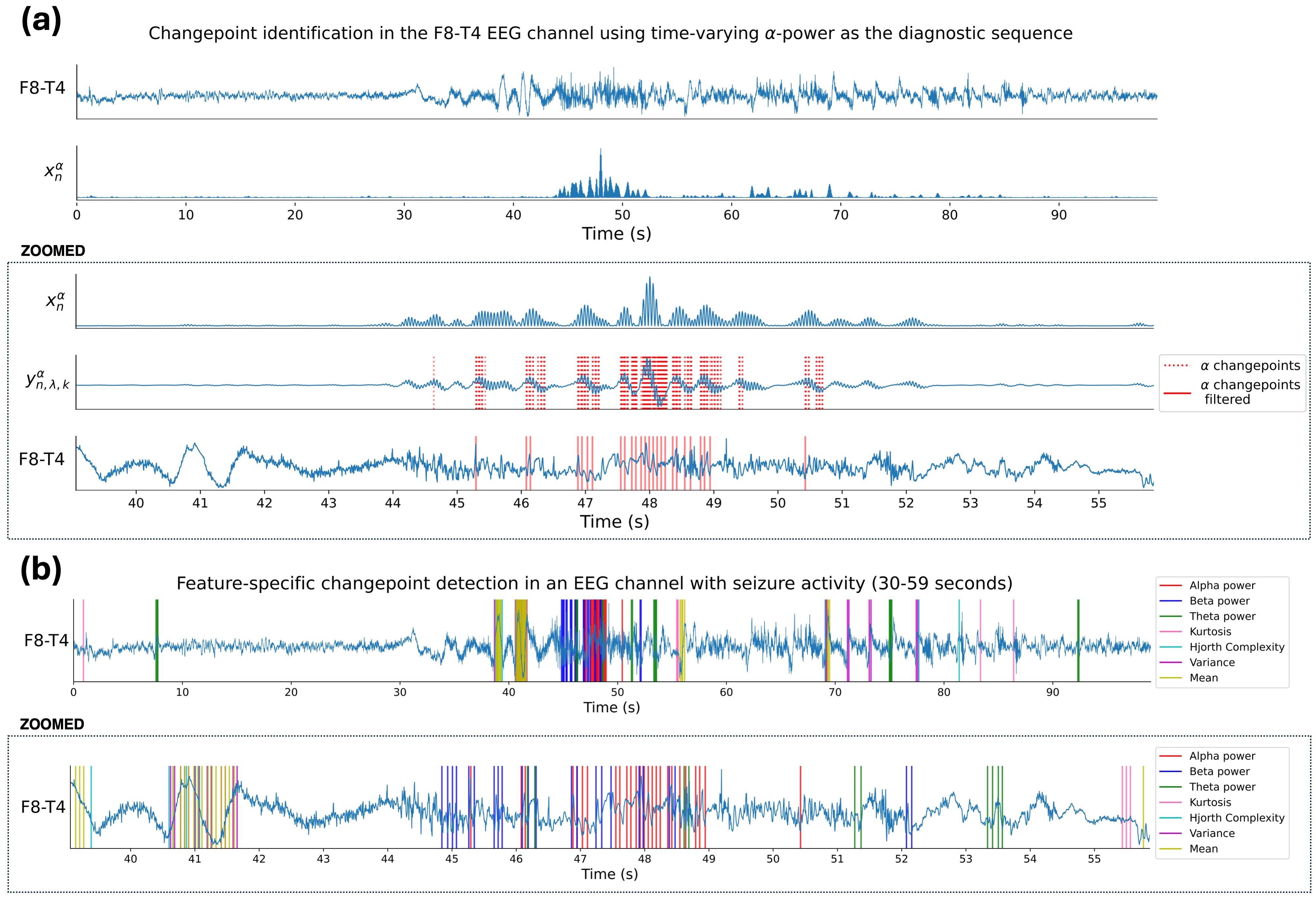}  % Replace with your PNG or PDF file
    
    \caption{Changepoint detection in nonstationary data. \textbf{(a)} The original F8-T4 EEG channel and the diagnostic sequence depicting the evolution of the $\alpha$-power as a function of time. Zoom-in on the section of high $\alpha$-power variations shows the value of the rolling difference $y^\alpha_{n,\lambda,\kappa}$, and the points outside of the 4 standard deviations threshold (dotted red lines). Finally, density threshold within a period equivalent to the lag determines the sub-segments of sustained variations in the original signal (solid red lines). \textbf{(b)} The F8-T4 EEG channel and various feature-specific changepoints. Observe that the changepoints are indicative of either a general but smoother transition into another state, or of a sharp signal variation. The latter can be seen by the identification of changepoints between 68 and 85 seconds, where each peak (indicative of a seizure), is identified as a changepoint. The changepoints used to segment the EEG before data augmentation are filtered once more, to remove any changepoints less than 256 samples (1 s) apart.}
    \label{fig:cp}
\end{figure*}

EEG signals exhibit temporally evolving signal statistics, due to variations in brain activity and/or the presence of artefacts \cite{2, 3}. This makes feature analysis a very challenging task in the characterisation and classification of EEG data. To this end, inspired by the work of Brodsky \textit{et al.} \cite{brodsky}, a non-parametric offline changepoint detection method was devised based on the evolution of a set of features in the EEG signal. The work in \cite{brodsky} shows that detecting changes in a signal can be simplified to detecting changes in the expected value of an ``engineered sequence'' derived from the original data stream, termed a \textit{diagnostic sequence} $(x^f_n)_{n \geq 1}$, where $f$ indexes the chosen feature. The task then boils down to selecting a subset of features which are known to discriminate well between seizure and non-seizure EEG segments and are also able to vary during a seizure. Each of such features corresponds to a different diagnostic sequence. This is desirable, as performing changepoint detection via diagnostic sequence provides additional control over the kind of dynamical regime changes that are detected. Our chosen EEG features are the power in the $\theta$ (4-8 Hz), $\alpha$ (8-12 Hz), and $\beta$ (12-30 Hz) frequency bands; Hjorth complexity \cite{hjorthcomp}; variance; mean and kurtosis. Each frequency band was extracted using a 5\textsuperscript{th}-order Butterworth bandpass filter and its output was squared, as in \cite{brodsky}. A 64-sample long rolling window (fs = 256 Hz) with a stride of 1 was used to obtain the Hjorth complexity, kurtosis, mean and variance diagnostic sequences. Figure \ref{fig:cp}(a) depicts an EEG trace along with its squared $\alpha$-power as a function of time. 

After computing these feature-specific diagnostic sequences, the changepoint locations are estimated by first defining for each feature, $f$, a sequence of exponentially weighted averages, $(x^f_{n,\lambda})_{n \geq 1}$, with \begin{equation} x^f_{n,\lambda} := \sum_{\ell = 1}^n \lambda^{n-\ell} x^f_\ell / w_{n,\lambda} \end{equation}where $w_{n,\lambda} := \sum_{\ell=1}^n \lambda^{n - \ell}$, and $\lambda$ is a \textit{forgetting factor}. Each $(x^f_{n,\lambda})_{n \geq 1}$ was then used to define a sequence of lagged differences $(y^f_{n,\lambda,\kappa})_{n\geq \kappa + 1}$ with \begin{equation}y^f_{n,\lambda,\kappa} := x^f_{n,\lambda} - x^f_{n-\kappa,\lambda}\end{equation} where $\kappa \geq 1$ denotes the lag. Then, the sequence $(y^f_{n,\lambda,\kappa})_{n\geq \kappa + 1}$ is used for changepoint detection for a feature $f$. The downweighting of historic observations provides a smoothing that increases resistance to false positives, while the choice of a lag ensures robustness against outliers, as a change to the EEG due to a seizure is likely to be sustained.\\

A forgetting factor, $\lambda$, of 0.9 and a lag, $\kappa$, of 16 samples (62.5 ms, given fs = 256 Hz) were chosen. Any significant changes in the diagnostic sequence should be reflected by a sharp peak in $(y^f_{n,\lambda,\kappa})_{n\geq \kappa + 1}$. These peaks were obtained by finding all the indices $i^*$ that provide a value $y^f_{i^*,\lambda,\kappa}$ greater than 4 standard deviations from the (unweighted) mean of $(y^f_{n,\lambda,\kappa})_{n\geq \kappa + 1}$ over the full observation window, as shown by the dotted red lines in Figure \ref{fig:cp}(a). An index $i^*$ is considered a changepoint if at least 70\% of the $i\in\mathbb{N}$ which satisfy $\lfloor i^*-\kappa/2 \rfloor \leq i \leq \lceil i^* + \kappa/2\rceil$ correspond to values $y_{i,\lambda,\kappa}^f$ above the 4 standard deviations threshold (solid red lines Figure \ref{fig:cp}(a)). This ensures that only significant changes in the data, lasting at least as long as the chosen lag, are selected. 

Finally, the feature-specific changepoints (Fig. \ref{fig:cp}(b)) are combined and filtered, such that they are separated by at least 256 samples (1 second at 256 Hz). The so obtained changepoints are shown in Figure \ref{fig:example}, and are used to segment the EEG data into smaller windows. These are augmented individually before their concatenation to create a full surrogate data sample. 

\subsection{Changepoint detection in ECG}\label{ecg_peak_detection_section}

The ECG peak detection algorithm is applied to the ECG and negated ECG signal to identify both non-inverted and inverted R peaks in the ECG signal. We use SciPy's \textit{find\_peaks} function for peak detection. To ensure comprehensive detection in cases of low signal quality or missing peaks, the algorithm implements an adaptive gap-filling mechanism. When intervals between detected peaks exceed a predefined maximum threshold, additional fixed points are added - points where the value in the surrogate segment is fixed to be the same as the original. These fixed points are interpolated to have equidistant gaps no greater than the maximum threshold. This hybrid approach ensures robust peak detection whilst maintaining physiologically plausible peak spacing. Two distinct configurations were evaluated; using a minimum peak distance of 50 and 60 samples (166.7 ms and 200 ms) and maximum interval thresholds of 150 and 80 samples (500 ms and 266.7 ms), respectively. Reducing the minimum peak spacing and maximum interval thresholds increases morphological similarity between the original and surrogate data as it increases the number of preserved points in the surrogates.

\subsection{Fixed edges iAAFT for EEG augmentation}

Following EEG changepoint detection, we apply a modified version of the iAAFT algorithm to the so determined quasi-stationary sub-segments. The iAAFT algorithm generates surrogate time series by initially shuffling the original data, and iteratively refining the power spectrum and histogram of the surrogate to converge to those of the original data. Our modified version of the iAAFT algorithm ensures that when concatenating the synthetic EEG sub-segments together, no peaks are created or removed which were or were not present in the original EEG segment. To do this, the edges in the sub-segment surrogates are forced to be the same as the original sub-segment, as designated by the highlighted red traces in Figure \ref{fig:example}. In our experiments, the number of fixed samples at the edges was 10\% of the sub-segment length. The modified iAAFT algorithm is outlined in Algorithm \ref{alg:iAAFT}.

\begin{algorithm}
\caption{\hspace{-0.1cm}\textbf{.} Fixed edges iAAFT}\label{alg:iAAFT}
\footnotesize		
\begin{algorithmic}
    \State \textbf{Input:} X is a 1D time series, Ns is the number of surrogates to produce, M is the proportion of the segment length used as fixed edges.
    \State \textbf{Output:} Xs is a 2D array of shape (N, Ns) containing the surrogates. 
    \\
    \State \textbf{Initialise} iter = 0, max\_it = 1000, threshold = $10^{-6}$, MSE\_start = 100, MSE = 1000, N = length of X, margin = M$\times$N, $X_{start} = X[:$margin$]$, $X_{end} = X[-$margin$:]$
    \\
    \If{X is real}
        \State \textbf{Initialize} $Xs \gets$ zeros(N, Ns)
        \State $X(\omega)$ $\gets$ fft(X) 
        \State $A \gets  |X(\omega)|$
        \State $X\_sorted \gets$ sorted(X)
        \State $X = X[margin:-margin]$
        \For{$a \gets 1$ to Ns}
            \State $r \gets $shuffle$(X)$
            \State $r = concatenate(X_{start}, r, X_{end})$
            \State MSE\_prev $\gets$ MSE\_start
            \While{($|\text{MSE} - \text{MSE\_prev}| > \text{threshold)}$ and (iter $<$ max\_it)}
                \State MSE\_prev $\gets$ MSE
                \State $\theta \gets \angle $ fft($r$)
                \State $s \gets$ ifft($A e^{\theta\text{i}})$
                \State Ind $\gets$ indices of sorted $s$
                \State $r[\text{Ind}] \gets X\_sorted$
                \State $r = s[margin:-margin]$
                \State $r = concatenate(X_{start}, r, X_{end})$
                        
                \State MSE $\gets$ MSE($A, |$fft($r$)$|$)
                \State iter $\gets$ iter + 1
            \EndWhile
            \State $Xs[:, a] \gets r$
            \State iter $\gets 0$      \EndFor
    \EndIf
    \State \Return $Xs$
\end{algorithmic}
\end{algorithm}

\subsection{Fixed points iAAFT for ECG augmentation}\label{iaaft_ecg_section}

Following peak detection, we apply a modified version of the iAAFT algorithm to the ECG data. As we do not segment ECG into smaller sub-segments, the edges of the ECG trace do not need to be fixed. Rather, we use the identified peaks and \(M\) of their neighbouring points as fixed points. To do this, the value of the ECG signal at the indices of the peaks and their \(M\) neighbours are replaced by the original signal values at each iteration of the iAAFT. This preservation margin of \(M\) samples on either side of each detected peak controls the extent of peak morphology preservation — larger values maintain more of the original signal around each peak, resulting in surrogates more closely resembling the original data. This version of the modified iAAFT algorithm is outlined in Algorithm 2. Finally, a Gaussian filter with standard deviation (\(\sigma=5\)) was applied to the data excluding the detected peaks, with interpolation used to fill the resultant gaps.

\begin{algorithm}
\caption{\hspace{-0.1cm}\textbf{.} Fixed points iAAFT}\label{alg:iAAFT_fixed_points}
\footnotesize		
\begin{algorithmic}
    \State \textbf{Input:} X is a 1D time series, Ns is the number of surrogates to produce, fixed\_indices is a list of indices to keep fixed, M is the number of points to fix around each fixed point
    \State \textbf{Output:} Xs is a 2D array of shape (N, Ns) containing the surrogates
    \\
    \State \textbf{Initialise} iter = 0, max\_it = 3000, threshold = $10^{-6}$, MSE\_start = 100, MSE = 1000, N = length of X
    \State expanded\_indices $\gets$ empty list
    \For{index in fixed\_indices}
        \State start\_margin $\gets$ max(0, index - margin)
        \State end\_margin $\gets$ min(N, index + margin + 1)
        \State margin\_indices $\gets$ [start\_margin, ..., end\_margin - 1]
        \State expanded\_indices $\gets$ expanded\_indices $\cup$ margin\_indices
    \EndFor
    \State fixed\_indices $\gets$ unique(expanded\_indices)
    \State fixed\_values $\gets X[$fixed\_indices$]$
    \\
    \If{X is real}
        \State \textbf{Initialize} $Xs \gets$ zeros(N, Ns)
        \State $X(\omega) \gets$ fft(X)
        \State $A \gets |X(\omega)|$
        \State $X\_sorted \gets$ sorted(X)
        \State non\_fixed\_mask $\gets$ ones(N, boolean)
        \State non\_fixed\_mask[fixed\_indices] $\gets$ False
        \State non\_fixed\_indices $\gets$ indices where value is not fixed
        \For{$a \gets 1$ to Ns}
            \State $r \gets$ copy(X)
            \State temp $\gets$ shuffle(length(X[non\_fixed\_mask]))
            \State $r[$non\_fixed\_mask$] \gets X[$non\_fixed\_mask$][$temp$]$
            \State MSE\_prev $\gets$ MSE\_start
            \While{($|\text{MSE} - \text{MSE\_prev}| > \text{threshold)}$ and (iter $<$ max\_it)}
                \State MSE\_prev $\gets$ MSE
                \State $\theta \gets \angle$ fft($r$)
                \State $s \gets$ real(ifft($A e^{\theta\text{i}}$))
                \State $s\_non\_fixed \gets s[$non\_fixed\_mask$]$
                \State $X\_sorted \gets$ sorted($X[$non\_fixed\_mask$]$)
                \State Ind $\gets$ argsort($s\_non\_fixed$)
                \State $r \gets$ copy($s$)
                \State $r[$non\_fixed\_mask$] \gets X\_sorted[$argsort(Ind)$]$
                \State $r[$fixed\_indices$] \gets$ fixed\_values
                \State MSE $\gets$ MSE(A,$|$fft($r$)$|$)
                \State iter $\gets$ iter + 1
            \EndWhile
            \State $Xs[:, a] \gets r$
            \State iter $\gets 0$
        \EndFor
    \EndIf
    \State \Return $Xs$
\end{algorithmic}
\end{algorithm}

\section{\bfseries Deep learning classification tasks}
\label{sec:3}

\subsection{Seizure detection task}

To validate the proposed data augmentation method, the performances of a deep learning seizure detection algorithm with and without data augmentation were compared using two public datasets. The well-known convolutional neural network EEGNet-8,2 \cite{eegnet} was used as a DL technique. In both the baseline and augmented experiments, we used the Adam optimiser with a learning rate of 0.001, a batch size of 32, 150 epochs, and a dropout rate of 0.25.\\

The public Children’s Hospital Boston - Massachusetts Institute of Technology (CHB-MIT) and Siena scalp EEG databases were selected for experimental validation. The CHB-MIT dataset contains EEG recordings from 23 subjects, aged between 1.5 and 22 years old who suffer from refractory epilepsy \cite{chbmit}. The EEG recordings all have a sampling frequency of 256 Hz and follow the longitudinal bipolar montage. The channels present in the recordings are:  Fp1-F7, F7-T3, T3-T5, T5-O1, Fp2-F8, F8-T4, T4-T6, T6-O2, Fp1-F3, F3-C3, C3-P3, P3-O1, Fp2-F4, F4-C4, C4-P4, P4-O2, Fz-Cz, Cz-Pz. The Siena scalp EEG database, contains recordings from 14 subjects, aged between 20 and 71 years old \cite{siena}. The EEG recordings were resampled to 256 Hz and the channels were combined to follow the same longitudinal bipolar montage as the CHB-MIT dataset. Both datasets underwent identical minimal preprocessing where a 60 Hz and 50 Hz notch filter was applied to remove power line interference in the CHB-MIT and Siena dataset respectively.

For every seizure segment of the CHB-MIT and Siena dataset, one changepoint-informed and one non-changepoint-informed iAAFT surrogate were generated. The original and surrogate EEG traces were segmented into 10-second non-overlapping segments. The z-score standardisation was applied individually to every channel, following segmentation. To obtain a balanced dataset, enough background segments were sampled to match the number of original seizure segments. A leave-one-patient-out cross-validation strategy was used and we report the average classification performance. When augmenting the training set, there were twice as many seizure segments as there were background segments (each seizure segment is augmented once). The imbalance in the training set was accounted for by a weighted loss function. 

\subsection{AF detection task}

The neural network architecture, detailed in Table \ref{ECGCNN}, presents a compact convolutional design for ECG classification. The network architecture consists of an initial convolution followed by two parallel pathways to capture multi-scale temporal features through different kernel sizes (15 and 31). The outputs of the parallel pathways are concatenated and processed through four sequential convolutional blocks, with progressive channel expansion. This hierarchical feature extraction is followed by a three-layer classifier for final classification. Each convolutional and linear layer incorporates batch normalisation and ReLU activation, with dropout layers (0.3) strategically placed to prevent overfitting. In both the baseline and augmented experiments, we used the Adam optimiser with a learning rate of 0.001, a weight decay of $5\times10^{-5}$, a batch size of 128, and 100 epochs.

\begin{table}[!h]
\scriptsize
\centering
    \begin{threeparttable}
    \caption{ECG CNN Architecture}
    \label{ECGCNN}
    \begin{tabular}{lll}
        \hline
        \textbf{Layer} & \textbf{Configuration} & \textbf{Output Shape} \\
        \hline
        Input & - & $(1, L)$ \\
        \hline
        \multicolumn{3}{l}{\textbf{Initial Processing}} \\
        Conv1D & $k=15$, $s=2$, $c=24$ & $(24, L/2)$ \\
        MaxPool1D & $k=2$, $s=2$ & $(24, L/4)$ \\
        \hline
        \multicolumn{3}{l}{\textbf{Parallel Pathways}} \\
        Path 1: Conv1D & $k=15$, $s=1$, $c=12$ & $(12, L/4)$ \\
        Path 2: Conv1D & $k=31$, $s=1$, $c=12$ & $(12, L/4)$ \\
        Concatenate & - & $(24, L/4)$ \\
        \hline
        \multicolumn{3}{l}{\textbf{Feature Extraction}} \\
        Block 1: Conv1D & $k=9$, $s=1$, $c=48$ & $(48, L/8)$ \\
        MaxPool1D & $k=2$, $s=2$ & $(48, L/16)$ \\
        Block 2: Conv1D & $k=9$, $s=1$, $c=96$ & $(96, L/16)$ \\
        MaxPool1D & $k=2$, $s=2$ & $(96, L/32)$ \\
        Block 3: Conv1D & $k=9$, $s=1$, $c=96$ & $(96, L/32)$ \\
        MaxPool1D & $k=2$, $s=2$ & $(96, L/64)$ \\
        Block 4: Conv1D & $k=5$, $s=1$, $c=192$ & $(192, 1)$ \\
        \hline
        \multicolumn{3}{l}{\textbf{Classifier}} \\
        Linear & $192 \rightarrow 96$ & $(96)$ \\
        Linear & $96 \rightarrow 48$ & $(48)$ \\
        Linear & $48 \rightarrow 2$ & $(2)$ \\
        \hline
    \end{tabular}
    \begin{tablenotes}
    \small
    \item $k$: kernel size, $s$: stride, $c$: output channels, $L$: input length. Each convolutional and linear layer was followed by BatchNorm and ReLU. Dropout (0.3) was applied after each block except the final layer
    \end{tablenotes}
    \end{threeparttable}
\end{table}

The ECG classification task was conducted using the Computing in Cardiology Challenge (CinC) 2017 dataset \cite{cinc}. The dataset comprises single-lead ECG recordings collected through AliveCor's handheld device. Each recording spans 9-61 seconds, sampled at 300 Hz with a 0.5-40 Hz bandwidth. The binary classification task used recordings labelled as either normal (5,076 samples) or those displaying atrial fibrillation (758 samples), following expert validation. Each atrial fibrillation (AF) segment underwent augmentation once using: (i) changepoint-informed iAAFT, (ii) non-changepoint-informed iAAFT, (iii) Gaussian random noise addition (\(\sigma=0.01\)), (iv) time-inversion, and (v) random cutout with a probability of 10\%. Following this, the original and surrogate ECG segments are segmented into 10 seconds non-overlapping segments and z-score standardisation is applied. 10-fold cross-validation was employed; to prevent data leakage we ensured that segments derived from the same sample were allocated to the same set.

\section{\bfseries{Results}}\label{resuls_section}

\begin{figure*}[!t]
    \centering
    \includegraphics[width=\textwidth]{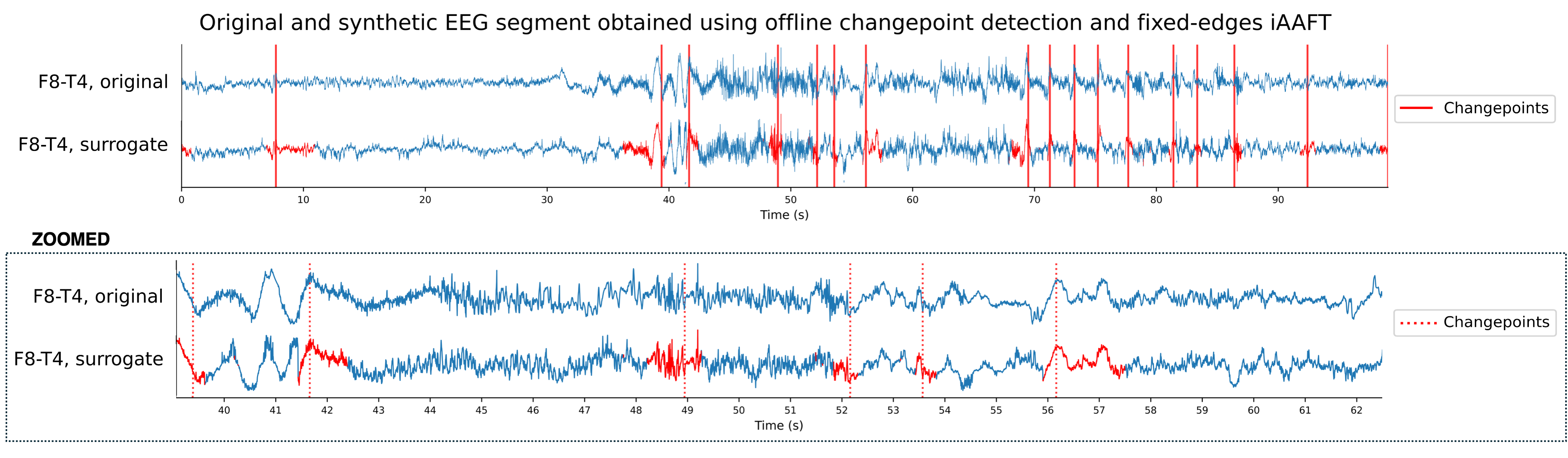}  % Replace with your PNG or PDF file
    \vspace{-0.6cm}
    \caption{Original EEG segment and augmented EEG segment. Note that around the changepoints, the edges of the surrogate are identical to the original data.}
    \label{fig:example}
    %\vspace{-0.2cm}
\end{figure*}

\begin{figure*}
    \centering    \includegraphics[width=0.9\textwidth]{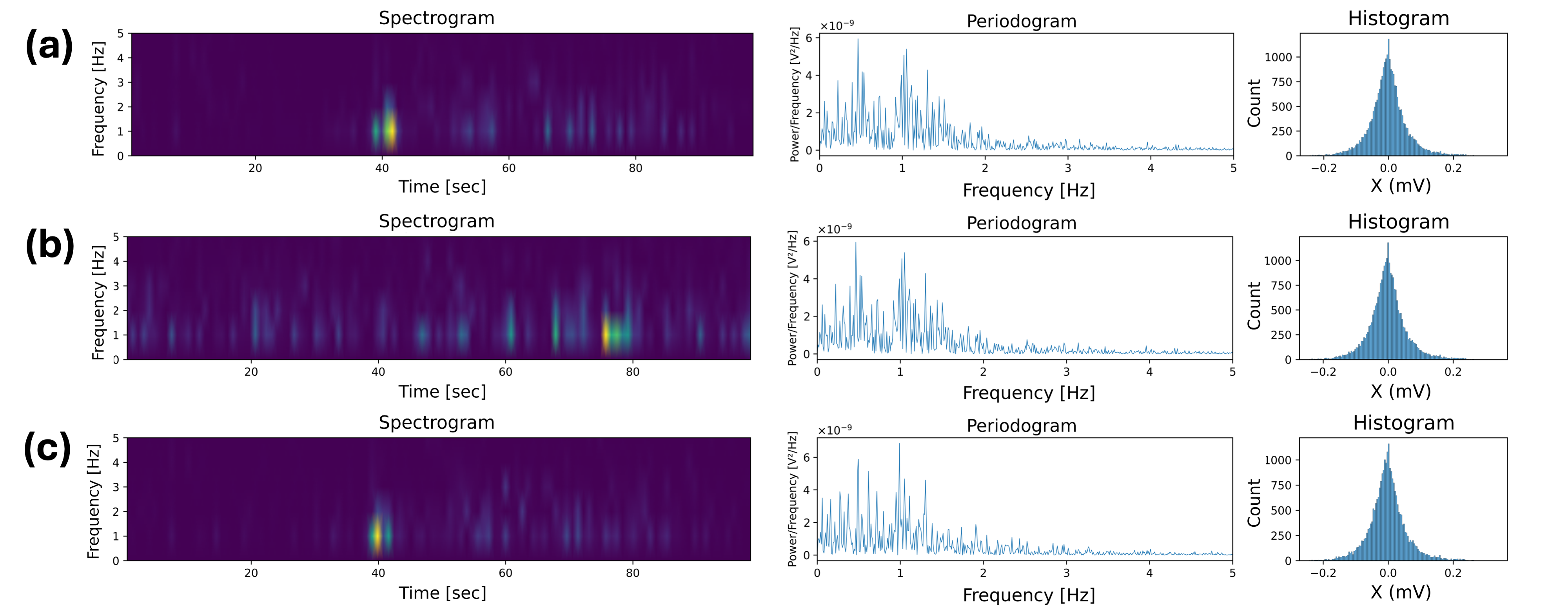}  % Replace with your PNG or PDF file
    \vspace{-0.2cm}
    \caption{The time-frequency spectrogram, power spectrum and histogram of the original F8-T4 EEG segment \textbf{(a)}, an iAAFT surrogate obtained without using offline changepoint detection \textbf{(b)}, and an iAAFT surrogate obtained using our offline changepoint detection method \textbf{(c)}. Although the power spectrum and histograms of both (a) and (b) are similar, the time-frequency characteristics of the original signal are lost in the standard surrogate. Our surrogate has a power spectrum that differs slightly from that of the original segment, but the overall time-frequency characteristics of the original signal are well preserved.}
    \label{fig:iaaft_nocp}
\end{figure*}

Figure \ref{fig:example} illustrates how the changepoint location and fixed edges (red traces) guide the EEG data augmentation. The time-frequency spectrogram, periodogram and histogram of the original signal, an iAAFT surrogate obtained without using our offline changepoint detection method, and an iAAFT surrogate obtained using our offline changepoint detection method are shown in Figure \ref{fig:iaaft_nocp}. The surrogate obtained using the proposed method was able to maintain the overall time-frequency characteristics of the original data. As the frequency of changepoints increased, the spectrogram of the surrogate converged to the spectrogram of the original signal. The sensitivity of our changepoint detection algorithm can be altered using the parameters described in Section \ref{changepointsection}: lag, forgetting factor, cutoff thresholds for changepoint filtering (4 standard deviations, and 70\%), and minimum distance between changepoints. \\

Table \ref{table:results} shows the results for the binary classification of seizure and background segments based on EEGNet-8,2. Observe an increase across all performance metrics when using the proposed method for both EEG datasets, with precision increasing by 10\% and 5.3\% for the CHB-MIT and Siena datasets respectively, indicating a significant drop in the number of false positives. In this case, the iAAFT augmentation without using changepoint detection resulted in slightly better performance than that based on the original data, but its success depends on the quality of the labels and the quasi-stationary nature of the data, neither of which can be relied upon. This confirms that changepoint-informed surrogates retain the characteristics of the original seizure segments, and are good candidates for the automated augmentation of nonstationary EEG data. 

\begin{table}[h]
    \centering
    \caption{Classification results on EEG datasets}
    \resizebox{\columnwidth}{!}{%
    \begin{tabular}{lcccc}
    \toprule % Thick top line
    \textbf{Model} & \textbf{Accuracy (\%)} & \textbf{Precision (\%)} & \textbf{Recall (\%)} & \textbf{F1-Score (\%)} \\ 
    \midrule % Regular line
    \multicolumn{5}{c}{\textbf{CHB-MIT Dataset}} \\
    \midrule
    Baseline & 68.2 & 68.3 & 63.0 & 61.7 \\
    iAAFT & 69.4 & 68.4 & 63.4 & 62.7 \\
    Ours & \textbf{72.6} & \textbf{78.3} & \textbf{66.6} & \textbf{66.5} \\ 
    \midrule
    \multicolumn{5}{c}{\textbf{Siena}} \\
    \midrule
    Baseline & 68.4& 70.3& 64.6& 65.6\\
    iAAFT & 66.3& 69.5& 63.5& 65.4\\
    Ours & \textbf{70.3}& \textbf{75.8}& \textbf{65.5}& \textbf{67.0}\\ 
    \bottomrule % Thick bottom line
    \end{tabular}
    }
    \label{table:results}
\end{table}

\begin{figure*}
    \centering
    \includegraphics[width=0.96\textwidth]{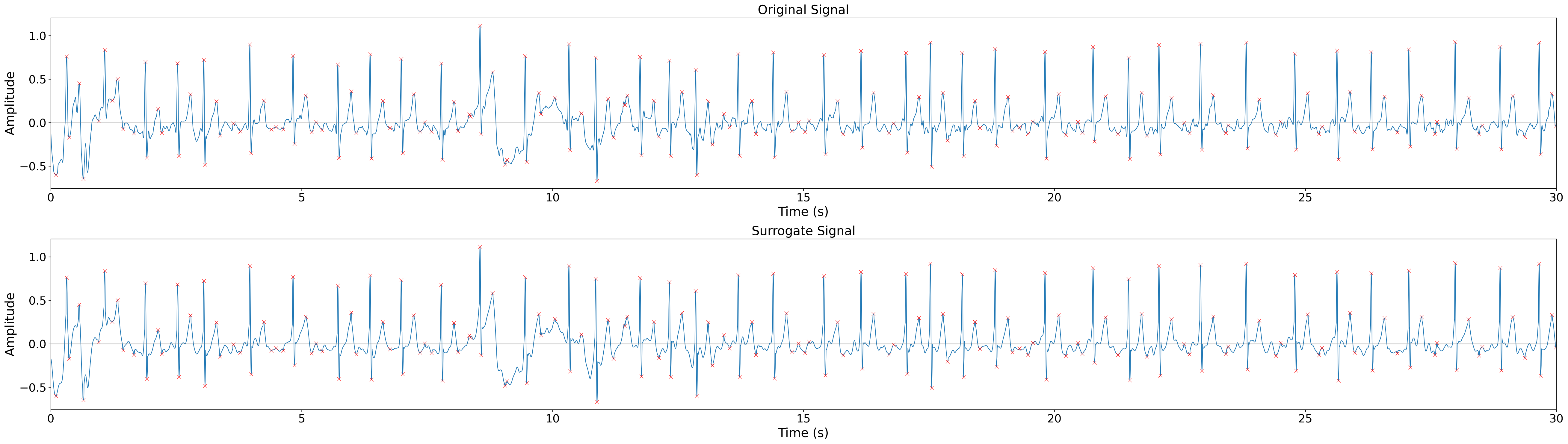}  
    \vspace{-0.25cm}
    \caption{Original ECG segment (top) and augmented ECG segment (bottom) obtained using a minimum peak distance of 60 samples, a maximum peak distance of 80 samples, and a margin \(M\) of 10 samples (hyperparameter configuration B - detailed in Section \ref{resuls_section}). Note that the amplitude and location of the detected peaks, denoted as red crosses, are identical in the surrogate and the original data.}
    \label{fig:ECG_augmentation}
    \vspace{-0.35cm}
\end{figure*}

Figure \ref{fig:ECG_augmentation} illustrates how the location of the prominent peaks, denoted by red crosses, for an ECG segment displaying AF, guides the data augmentation process. Whilst the surrogate signal preserves both the temporal position and amplitude of these peaks, it introduces controlled variability in the ECG waveform morphology. The degree to which the original signal morphology is retained is governed by several parameters: the maximum interval threshold and minimum peak distance, as detailed in Section \ref{ecg_peak_detection_section}, alongside the margin \(M\) and the standard deviation of the Gaussian filter, which are described in Section \ref{iaaft_ecg_section}. We evaluated our method using two distinct hyperparameter configurations. Configuration A utilises a minimum peak distance of 50 samples, a maximum peak distance of 150 samples, and a margin \(M\) of 5 samples. Configuration B utilises a minimum peak distance of 60 samples, a maximum peak distance of 80 samples, and a margin \(M\) of 10 samples. \\

The performance metrics for the AF classification task using our ECG CNN architecture (detailed in Table \ref{ECGCNN}) are reported in Table \ref{table:results_ecg}. The proposed method demonstrates superior performance compared to the majority of baseline approaches, apart from a slight decrease in recall in comparison to the time-inversion augmentation method.

\begin{table}[h]
    \centering
    \caption{Classification results using the ECG CinC dataset}
    \resizebox{\columnwidth}{!}{%
    \begin{tabular}{lcccc}
    \toprule % Thick top line
    \textbf{Model} & \textbf{Accuracy (\%)} & \textbf{Precision (\%)} & \textbf{Recall (\%)} & \textbf{F1-Score (\%)} \\ 
    \midrule % Regular line
    Baseline & 97.0 & 62.1 & 75.3 & 66.9 \\
    Noise & 97.1 & 63.9 & 72.5 & 67.0 \\
    iAAFT & 97.2 & 62.3 & 75.5 & 68.0 \\
    Time-inversion & 97.0 & 59.7 & \textbf{79.4} & 67.8 \\
    Cutout & 97.3 & 65.9 & 72.6 & 68.5  \\
    Ours (A) & 97.3 & 64.2 & 76.1 & 69.0  \\ 
    %Ours 2 & 97.3 & 65.3 & 70.6 & 67.3  \\ 
    %Ours 14 & 97.3 & 65.9 & 75.0 & 68.9  \\ 
    %Ours 15 & 97.0 & 61.3 & 75.4 & 66.6  \\ 
    Ours (B) & \textbf{97.5} & \textbf{68.1} & 75.4 & \textbf{69.7}  \\ 
    \bottomrule % Thick bottom line
    \end{tabular}
    }
    \label{table:results_ecg}
    \vspace{-0.3cm}
\end{table}

\section{\bfseries Discussion}

We have introduced a novel data augmentation technique suited to nonstationary data. This has been achieved by first identifying changepoints in signal dynamics by using a changepoint detection algorithm which relies on multiple feature-specific diagnostic sequences for EEG and peak-detection in ECG. For EEG data, the proposed modified iAAFT algorithm has been applied to the sub-segments determined by the changepoints. Finally, the synthetic sub-segments are concatenated, creating a full surrogate EEG sample. For ECG data, the modified iAAFT algorithm has been applied to the entire signal whilst maintaining the position and amplitude of the detected peaks from the original signal. Through real-world EEG-based seizure detection and ECG-based AF detection, we have shown that models trained using the proposed data augmentation method have outperformed those trained without data augmentation, and those trained with common augmentation techniques. \\

Our method is straightforward, interpretable and does not require extensive training data. While the improvement in performance for AF detection is less significant compared to seizure detection, this stems primarily from constraints in peak detection reliability. To handle cases where peak detection was less reliable, we had to set the minimum time between fixed points to be 166.7 ms and 200 ms in configuration A and B respectively. While this ensures that signal characteristics are preserved, it restricts the diversity of generated surrogates. Though initially considered, the Pan-Tompkins algorithm \cite{pantompkins} displayed inconsistent peak detection performance across samples and was therefore not implemented. A more robust PQRST detection algorithm would enable greater precision in fixed point placement, potentially enhancing the method's effectiveness by allowing greater variability in ECG surrogates. 

%\clearpage

\end{document}